\documentclass[pdflatex,sn-mathphys-num]{sn-jnl}

\usepackage{graphicx}%
\usepackage{multirow}%
\usepackage{amsmath,amssymb,amsfonts}%
\usepackage{amsthm}%
\usepackage{mathrsfs}%
\usepackage[title]{appendix}%
\usepackage{xcolor}%
\usepackage{textcomp}%
\usepackage{manyfoot}%
\usepackage{booktabs}%
\usepackage{algorithm}%
\usepackage{algorithmicx}%
\usepackage{algpseudocode}%
\usepackage{listings}%

\theoremstyle{thmstyleone}%
\theoremstyle{thmstyletwo}%

\theoremstyle{thmstylethree}%

\DeclareMathOperator*{\argmin}{arg\,min}
\begin{document}

\title[Article Title]{A 3D hybrid cellular Potts model with a discrete deformable fiber network: modeling cell contraction and extracellular matrix remodeling}


\author[1]{\fnm{Koen A.E.} \sur{Keijzer}}
\email{k.a.e.keijzer@math.leidenuniv.nl}

\author*[1,2]{\fnm{Roeland M.H.} \sur{Merks}}
\email{merksrmh@math.leidenuniv.nl}

\affil[1]{\orgdiv{Mathematical Institute}, \orgname{Leiden University}, \orgaddress{\street{Einsteinweg 55}, \city{Leiden}, \postcode{2333 CC},  \country{The Netherlands}}}

\affil[2]{\orgdiv{Institute of Biology}, 
\orgname{Leiden University}, \orgaddress{\street{Einsteinweg 55}, \city{Leiden}, \postcode{2333 CC}, \country{The Netherlands}}}



\abstract{
The extracellular matrix (ECM) is a fibrous and dynamic network that plays a critical role in development, homeostasis, and disease.
Cells both respond to and remodel the ECM, engaging in a mechanical reciprocity that shapes tissues.
To study these interactions, computational models have been developed that simulate either ECM mechanics or cell behavior.
The Cellular Potts Model (CPM) is a flexible cell-based framework that has been extended to many biological processes, including coupling to a discrete deformable fiber network.
So far, however, this extension has only been applied in two dimensions, even though a three-dimensional setting is biologically more relevant.
Here, we present a 3D hybrid framework that couples the CPM with a discrete and deformable representation of the ECM. This model enables explicit simulation of cell-induced ECM remodeling, including fiber reorientation and matrix densification. By incorporating contractile forces through static adhesion points, the model captures how ECM elasticity and fiber stiffness influences cell shape. The simulations show that ECM stiffness, controlled by fiber crosslinking, resists contraction and reaches equilibrium, while crosslink density modulates fiber alignment and local matrix accumulation. This framework provides a versatile platform for studying cell–ECM mechanics and supports future studies of multicellular behavior in realistic 3D environments.
}




\maketitle

\section{Introduction}
The extracellular matrix (ECM) plays a central role in development, tissue homeostasis, and disease. It contributes to processes such as cancer invasion, tissue morphogenesis, and angiogenesis~\citep{heisenbergForces2013,kimECMintegrin2022,duThreeDimensional2016}. The ECM is a network of fibrous proteins including collagen, fibronectin, and laminin~\citep{rozarioExtracellular2010}, which provides structural support and mediates biochemical and mechanical signals. For example, vascular endothelial growth factor (VEGF) can bind to the ECM and later be sensed by endothelial cells as they degrade the ECM using matrix metalloproteinases (MMPs)~\citep{parkVascular1993,leeProcessing2005,chenAnchorage2010}. Cells respond to mechanical features of the ECM, such as stiffness, which they can sense in their surroundings. For example, different cells, such as endothelial cells and fibroblasts, spread out more on stiff substrates than on soft ones~\citep{reinhart-kingDynamics2005,grolmanExtracellular2020}. In addition to reacting to local stiffness, cells can detect spatial variations in stiffness and migrate upwards or downwards these stiffness gradients, in a process known as durotaxis~\citep{mathieuPositive2024,senguptaPrinciples2021}. Furthermore, the fiber orientation of the ECM can be sensed by cells. This can be a static response, where cells change their morphology based on the surrounding tissue~\citep{friedrichsCellular2007}, or a dynamic one, where cells migrate more easily along aligned fibers~\citep{wangExtracellular2018}, a process called topotaxis or contact guidance~\citep{senguptaPrinciples2021}. Cells not only respond to the ECM but also actively engage with it. They can degrade ECM with MMPs, deposit new matrix~\citep{jenkinsLaminin2016}, stiffen the ECM~\citep{karlinskizurAltered2025}, and by exerting contractile forces, induce local changes in ECM density~\citep{malandrinoDynamic2019} and reorient fibers. This mutual mechanical interaction between cells and the ECM, in which cells both sense and modify their environment, is referred to as mechanical reciprocity~\citep{vanhelvertMechanoreciprocity2018}.

This mechanical interplay between cells and the ECM has inspired the development of a wide range of computational models. These models aim to investigate how cell behavior and matrix structure co-evolve through mechanical interaction. We focus specifically on models that include fibrous ECM structure and allow for the mechanical influence of cells on the ECM, with or without feedback from the ECM to the cell. We distinguish between models that consider (i) matrix remodeling due to cellular forces and (ii) changes in cell behavior due to ECM mechanical properties. Some recent models aim to combine both to capture full mechanical reciprocity.

We first review models that primarily focus on ECM deformation by cellular contraction (i).
Often, the ECM is modeled with finite element (FE) simulation software, where fibers are represented as single elements, or collections of elements and cellular contraction is modeled with a changing boundary condition~\citep{sopherNonlinear2018,sarkarUnexpected2024,gorenElastic2020,slaterTransient2021}.
These models are used to study the effect of fiber reorientation, buckling, and non-linear mechanics on the long-range propagation of the cellular contractile forces.
One finite element (FE) model~\citep{aghvamiFiber2016} compared the mechanical response of a continuous ECM with that of two fibrous ECMs differing in connectivity. This work showed that explicitly representing fibers enables long-range force transmission, and that fiber displacements in networks with lower connectivity contribute to this effect.
In a different study, rheology experiments are combined with a finite element model and they find that fiber buckling between two contractile cells can lead to a local softening of the tissue~\citep{sarkarUnexpected2024}. 
These models successfully describe how contractile forces align and stiffen the ECM, but simplify the cell shape as a circular contractile ring.

Another line of work emphasizes the effect of ECM mechanics on cell behavior (ii). For example, the PhysiCell software, a modeling framework for agent-based biological models, has extensions that allow cells to sense local fiber directions, densities and isotropy and also realign fibers by moving over them~\citep{dallonMathematical1999,noelPhysiMeSS2024, metzcarSimple2025} and has recently been used to study the effect of ribose-induced crosslinking on cancer invasion~\citep{botticelliHybrid2025}. This modeling approach is valuable for large scale models, however the phenomenological description of cell-ECM interactions, and the simplified ECM mechanics limit its use for smaller scale simulations where cell shape and single cell dynamics are important.

There are models that aim to capture both (i) and (ii) and attempt to capture full mechanical cell–ECM reciprocity. A 2D model in which both the cell and the fibers are discretized as individual agents was used to study non-linear fibrous ECM and dynamic cell shape, showing how durotaxis and co-alignment of cells can arise \citep{reinhardtComplex2013}.
Such a model begins to address the complex interplay between ECM architecture and cell dynamics in 2D. However, the interaction of a cell with ECM in 3D is quite different from that in 2D systems~\citep{doyleOnedimensional2009,doyleMechanosensing2016}, since, unlike 2D systems, there is no inherent downward polarization of the cell and steric constraints from surrounding fibers play a more prominent role.
Models that aim to take into account both ECM deformation and the cellular response to ECM stiffness and isotropy in 3D include, for example, the work of~\citep{eichingerComputational2021,pauknerWhat2023}. They use 3D FE models to model fibrous ECM and have point particles interact with them using mechanosensitive adhesion. Other work in this direction includes a study where both the ECM and the cell boundary are constructed from discrete elements, and the model is used to investigate filopodia formation \citep{kimComputational2022}.

While these approaches successfully incorporate detailed mechanics and 3D cell shapes, they differ from our work in their representation of cell dynamics. Our model builds upon the cellular Potts modeling (CPM) framework, which offers a well-established and extensible method for simulating cellular systems~\citep{granerSimulation1992,hirashimaCellular2017}.  There are already hybrid CPM models that extend the CPM framework with ECM models, such as the CPM-FEM models, that simulate cell-ECM interactions on the scale of subcellular protrusions and use a finite element simulation for the strain computations on a two-dimensional domain. These models are used to study how endothelial cells can form a network solely based on mechanical interactions~\citep{oersMechanical2014}, and how durotaxis and stiffness-dependent spreading emerge from mechanosensitive adhesion dynamics~\citep{rensCell2020,novikovaContractile2013}. Further work replaced the homogeneous, continuous ECM considered in the CPM-FEM models with fibrous structure to simulate the heterogeneous and non-linear fibrous ECM mechanics in 2D~\citep{tsingosHybrid2023,keijzerHow2025}. 

In this work, we extend the coupling between the CPM and fibrous ECM to 3D. We couple a 3D cellular Potts (CPM) model to a fully 3D bead-spring representation of the ECM, allowing contractile cells to deform and align fibers in a mechanically heterogeneous matrix. This model enables simulation of cell-driven ECM remodeling in 3D environments and captures long-range matrix deformation and anisotropy emerging from cellular contractility.
Unlike vertex or triangulated approaches, this coupling of the CPM framework to the fibrous ECM model allows for 
integration of existing CPM extensions such as chemotaxis~\citep{merksCell2006,dal-castelCompuCell3D2025}, cell division, cell polarization~\citep{beltmanLymph2007,niculescuCrawling2015}, and
tissue elongation~\citep{vroomansSegmentSpecific2015},
making it a flexible base for biologically relevant model extensions.

This framework offers a new computational tool to study how collective cell behavior reshapes the ECM and how the mechanical structure of the ECM emerges from cellular forces. In the remainder of this paper, we describe the model formulation, present simulations of ECM alignment and deformation, and compare our results to related modeling efforts.

\section{Methods}

\subsection{Model definition}
We present a 3D hybrid CPM model that extends earlier work coupling the CPM to a discrete fibrous bead–spring representation of the ECM in 2D~\citep{tsingosHybrid2023}. The hybrid model consists of two components: the CPM and the ECM (Figure~\ref{fig:overview}(A)). We first describe each component separately, followed by the details of their coupling. A single timestep of the simulation consists of one CPM cycle, after which the ECM model is integrated to steady state. The two models are therefore executed alternately for a prescribed number of timesteps.


\subsubsection{The cellular Potts model}
The cellular Potts model is a stochastic lattice-based model that evolves the shape of the cell over time. We define the model on a lattice $\Lambda = [0,L]^3 \cap\mathbf Z^3$ with $L$ an integer. That is, $\Lambda$ consists of the integer points of a cube of side lengths $L$. The CPM is defined by its spin field $\sigma$ that assigns a non-negative integer $\sigma_t(p)$ to each $p$ in a lattice $\Lambda \subseteq \mathbf Z^3$. For each point $p \in \Lambda$, we define its neighbourhood as
\begin{equation}\label{eq:NB}
\begin{aligned}
    \text{NB}(p) = \Big\{ \, 
         p + \Delta \in \Lambda :\;
        \Delta \in \{-1,0,1\}^3,\;
        \exists i \;\text{such that}\; \Delta_i = 0
    \Big\} \setminus \{p\},
\end{aligned}
\end{equation}
where $p+\Delta$ denotes coordinate-wise addition.  
This neighbourhood $\text{NB}(p)$ consists of the $18$ closest lattice points to $p$.

The spin field is updated using an adapted Metropolis–Hastings algorithm sampling two adjacent sites $p, p' \in \Lambda$ with $p'\in\text{NB}(p)$ with different spin values and considering replacing the value $\sigma\left(p\right)$ with the value of $\sigma\left(p'\right)$. Such a sampled pair $(p, p')$ is an interface, or edge of the spin field. Each timestep of the CPM, referred to as a Monte Carlo Step (MCS), consists of a number of copy attempts equal to the number of cellular interfaces in the system.  The acceptance probability of a copy-attempt $P(\Delta H)$, given as $P(\Delta H) = \min\{1, \exp\left(-\Delta H/\mu\right)\}$ with $\mu$ a motility-parameter, depends on the change in energy $\Delta H$ that the copy-attempt introduces into the system and is defined as 
\begin{equation}\label{eq:Hamiltonian}
    \Delta H = H_\text{new} - H_\text{old} + \Delta H_\text{work}, 
\end{equation}
where $H_\text{old}$ and $H_\text{new}$ are the system's energy state before and after the copy attempt and $\Delta H_\text{work}$ is an active energy. The coupling to the ECM model is done in the work term $\Delta H_\text{work}$ and is defined in a later section.

The term $H_\text{new} - H_\text{old}$ in Equation~\eqref{eq:Hamiltonian} describes the change in energy of the body of the cell before and after a copy-attempt where the energy of the cell is given via
\begin{equation}\label{eq:HamiltonH}
    H = \lambda_\text{volume} \left(V - v\right)^2 + J\sum_{(p,p')} 1_{\sigma\left(p\right) \neq \sigma\left(p'\right) },
\end{equation}
where $\lambda_\text{volume}$ and $J$ are scaling parameters, $V$ is the target volume and $v = \sum_{p} 1_{\sigma\left(p\right) = 1}$ is the volume of the cell.

\subsubsection{Extracellular matrix}
The extracellular matrix is modeled as in previous work~\citep{tsingosHybrid2023,keijzerHow2025} with the addition of a third dimension. In this section we briefly recap the mechanics. The ECM is modeled as a fibrous network where each fiber is made out of particles that are linked together by springs. The positions of the particles in the fiber network are vectors in $\mathbf R^3$ and they are updated according to an overdamped Langevin equation
\begin{equation}\label{eq:langevin}
    \frac{\mathrm d }{\mathrm d\, t} \vec{x_i}(t) = F_i(t) + \xi_i,
\end{equation}
where $F_i(t)$ is the force on the $i$-th particle and $\xi_i$ is a small random force. Equation~\eqref{eq:langevin} is numerically integrated to steady state using the molecular dynamics (MD) package HOOMD-blue~\citep{andersonHOOMDblue2020}. 
We distinguish between two types of particles: \emph{free} and \emph{fixed}. Only the positions of free particles are updated according to Equation~\eqref{eq:langevin}. Particles that at the start of the simulation are located outside the simulated domain $\Omega = [0,L]^3$ are always fixed (Figure~\ref{fig:overview}(B)), while particles inside $\Omega$ are either free or linked to the cell, as described in the coupling section.

The fiber network is initialized in two steps. The first step is the placement of the fibers and the second step is crosslinking fibers that are close together. Each fiber $i$ consists of $N_\text{beads}$ particles with positions $x_{i,0}, \dots, x_{i,N_\text{beads}-1} \in \mathbf R^3$. Fiber initialization is done by uniformly sampling initial positions for the fibers $\vec{x}_{i,0} \in \Omega $ for $i=1,\dots N_\text{fibers}$ and uniformly sampling directions $\vec{v}_i \in S^2$ with the number of fibers $N_\text{fibers}$ a parameter. Then the next particle in fiber $i$ is put a distance $r_0$ in the direction $\vec{v}_i$ from $\vec{x}_{i,0}$, leading to the fiber particles having coordinates 
$$
    \vec{x}_{i,j} = \vec{x}_{i,0} + j r_0\vec{v}_i, 
$$
for $j = 1,\dots, N_\text{particles}$, with $N_\text{particles}$ the number of particles for each fiber.

Once the fibers are formed, we crosslink the fiber network. We do this by choosing a maximal crosslinking number $N_\text{cross,max}$ and a distance $r_\text{cross,max}$. Then, $N_\text{cross,max}$ pairs are sampled from all pairs of particle beads with a distance at most $r_\text{cross,max}$. Then all sampled beads are crosslinked. Note that the total number of crosslinks depends on the number of crossings that are a distance $r_\text{cross,max}$ away from each other.

\subsubsection{Model coupling}
Having defined the CPM and ECM models separately, we now describe their coupling. The two models are executed consecutively. During a CPM Monte Carlo step (MCS), all ECM particles remain fixed in place, except for adhesion particles that may be displaced through CPM copy attempts. Conversely, during integration of Equation~\eqref{eq:langevin}, the CPM configuration is held fixed.  

The CPM is defined on the lattice $\Lambda = \Omega \cap \mathbf Z^3$, while ECM particles have continuous coordinates in $\mathbf R^3$. Each lattice site $p \in \Lambda$ corresponds to a voxel of side length $1$ in $\Omega$, defined as the set of positions $\vec{v} \in \Omega$ with $p_{\vec{v}} = (\lfloor x \rfloor, \lfloor y \rfloor, \lfloor z \rfloor)$.  

\begin{figure}[h!]
    \centering
    \includegraphics[width=\linewidth]{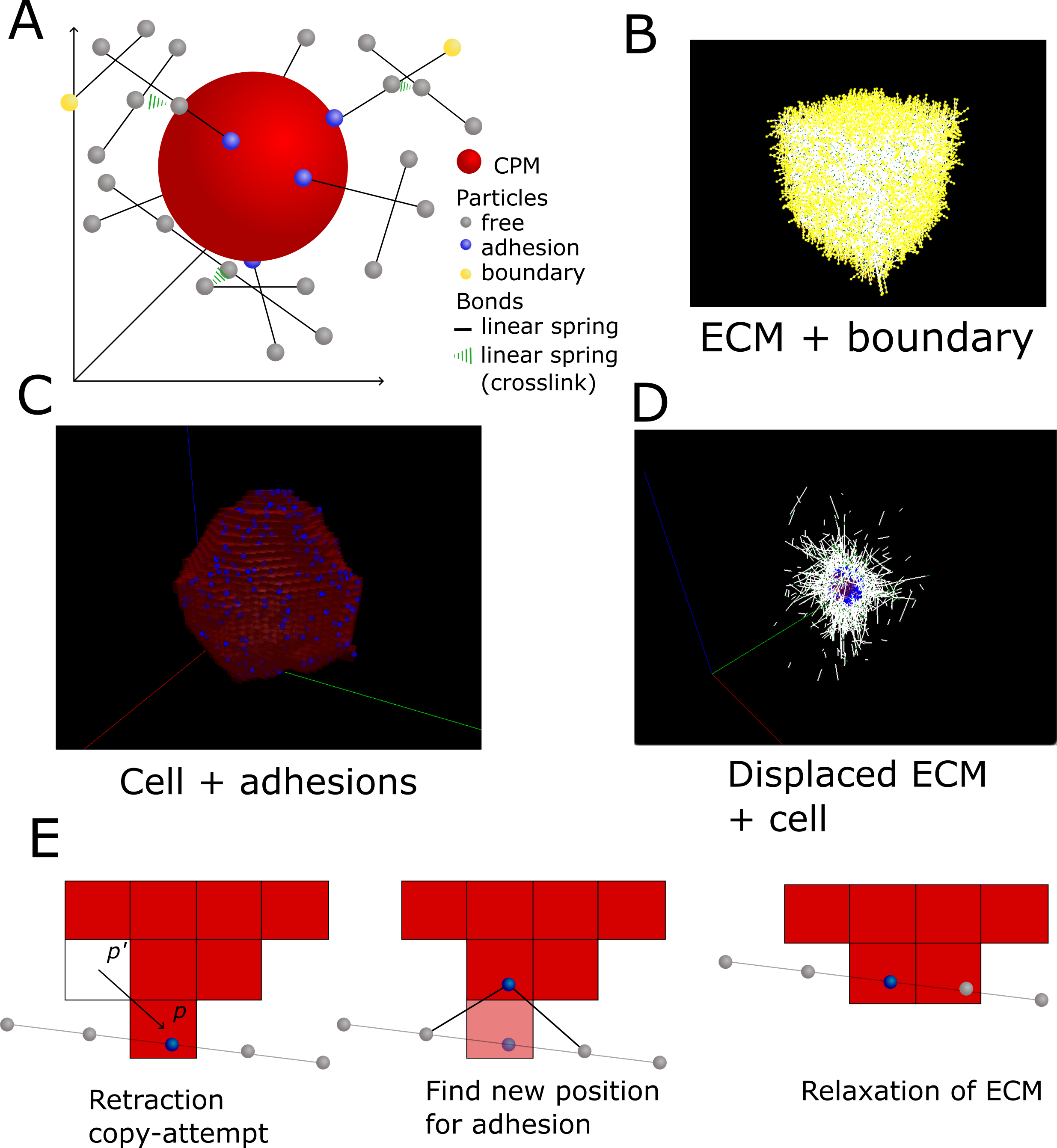}
    \caption{Overview of all model components. A: Sketch of CPM cell in MD ECM. B: Screenshot of a simulation with all fibers and yellow boundary particles drawn. Drawing all fibers is not informative. C: Screenshot where none of the fibers are drawn, only the adhesions. D: Contracted cell in ECM, all fibers that have been displaced at least $15\,\mu$m are drawn. E: Cartoon of the fundamental coupling between CPM and ECM. Cell attempts a copy attempt by replacing $\sigma(p)$ with $\sigma(p')$, for which it needs to displace an adhesion particle. After one MCS, the free particles in the MD simulation are updated, and the remainder of the ECM fiber is displaced.}
    \label{fig:overview}
\end{figure}

At initialization, a subset of ECM particles is designated as adhesion sites (Figure~\ref{fig:overview}(C)). These particles are of type `fixed': they are not updated by Equation~\eqref{eq:langevin} and remain stationary while the ECM relaxes to mechanical equilibrium. Instead, their displacement is governed entirely by the CPM, which moves the voxel to which the adhesion is attached (Figure~\ref{fig:overview}(D,E)). In this way, the CPM controls the motion of adhesions, and the ECM responds passively through deformation. The presence of an adhesion in a voxel $p$ contributes to the energy of copy attempts through the work term $\Delta H_\text{work}$ in Equation~\eqref{eq:Hamiltonian}.  

When the CPM sampling algorithm selects an edge $(p,p')$ with 
$p,p' \in \Lambda$, $p' \in \text{NB}(p)$, and $\sigma(p) \neq \sigma(p')$, a copy attempt may involve adhesion particles (Figure~\ref{fig:overview}(E)). In that case, $\Delta H_\text{work}$ accounts for the ECM forces associated with displacing those adhesions. To evaluate the contribution of moving an adhesion particle $\alpha$ by a displacement vector $\vec{v}$, we assume the ECM is in mechanical equilibrium and sum the contributions from the bonds directly attached to $\alpha$:  
\begin{align}
\Delta H_{\text{ECM},\alpha,\vec{v}} =
&\sum_{(\vec{x}_\alpha, \vec{y})} \frac{k}{2} \left[ (\| \vec{x}_\alpha - \vec{y}\| - r_0)^2  
- (\|\vec{x}_\alpha + \vec{v} - \vec{y}\| - r_0)^2 \right] \nonumber \\
&+ \sum_{\theta} \frac{t}{2} \left[ (\theta - \theta_0)^2 - (\theta_{\vec{v}} - \theta_0)^2 \right],
\label{eq:deltah:ecm}
\end{align}
where the first sum runs over harmonic bonds between the adhesion at $\vec{x}_\alpha$ and neighboring particles $\vec{y}$, with spring constant $k$ and rest length $r_0$, and the second sum runs over angular constraints involving the adhesion, with rest angle $\theta_0$, stiffness $t$ and resulting angle $\theta_{\vec{v}}$.  

During a copy attempt $(p,p')$, the work term $\Delta H_\text{work}$ may be 
influenced by adhesions located in either of the two voxels. To describe these 
cases explicitly, let $A_p$ and $A_{p'}$ denote the sets of adhesion particles in voxels $p$ and 
$p'$ respectively. If $A_p$ is non-empty, the adhesions in $p$ retract to a 
neighboring voxel $q \in \text{NB}(p)$ that still belongs to the same cell, 
$\sigma(p) = \sigma(q)$. The displacement vector $\vec{v}$ of the adhesions is then chosen to minimize the total energy cost,  
\begin{equation*}
    \vec{v} = \argmin_{
    \substack{\vec{v}: \\\vec{v} = p - q\\q\in \text{NB}(p)\\
    \sigma(q) = \sigma(p)}} \sum_{\alpha \in A_p} 
    \Delta H_{\text{ECM}, \alpha, \vec{v}} .
\end{equation*}
If $A_{p'}$ is non-empty, their displacement is aligned with the direction of 
the copy attempt, given by the normalized vector 
$\widehat{p-p'} = (p-p')/\|p-p'\|$.  

Combining both cases, the total work term is defined as
\begin{equation}\label{eq:Hwork}
    \Delta H_\text{work} = 
    \sum_{\alpha \in A_p} \Delta H_{\text{ECM},\alpha,\vec{v}} 
    + \sum_{\alpha\in A_{p'}} \Delta H_{\text{ECM},\alpha,\widehat{p-p'}}.
\end{equation}

\subsubsection{Model initialization}
A 3D rectangular grid of $200\times200\times200$ micrometers was used as a simulation domain. Each simulation starts with a spherical cell situated in the middle of the simulation domain with initial size 144000 $\mu\text{m}^3$. Next, the adhesion particles are initialized by choosing a maximum number of adhesions $N_\text{adh,max}$ and a minimal distance from the centroid $r_\text{init,adh}$. At the start of the simulation, $N_\text{adh,max}$ particles are randomly sampled from all the particles $\vec{x} \in\Omega$ inside the cell, $\sigma\left(p_{\vec{x}}\right) = 1$, with distance to the cell centroid at least $r_\text{init,adh}$. The sampled particles are turned into adhesion particles, and all particles closer than $r_\text{init,adh}$ to the cell's centroid are removed from the simulation.

\subsubsection{Implementation}
Our previous 2D TST-MD framework~\citep{tsingosHybrid2023,keijzerHow2025} was 
implemented in the Tissue Simulation Toolkit (TST) and linked to a Python module that uses HOOMD-blue~\citep{andersonHOOMDblue2020} to integrate Equation~\eqref{eq:langevin}. The TST is a widely used software package for CPM studies~\cite{merksCellcentered2005,dejongShapes2024,vroomansEvolution2023,colizziEvolution2020}, but it is specialized for 2D simulations. For the present work, we therefore developed a custom 3D CPM implementation, which we coupled to the existing ECM code to form the hybrid CPM model. The ECM module was originally developed in collaboration with the eScience Center. This work was performed using the ALICE compute resources provided by Leiden University.

\subsection{Quantification of network remodeling}
To measure the effect of cell contraction on the fibrous ECM network we consider the realignment of the fibers in the network and the increase in density. 

\subsubsection{Realignment of the network}
To quantify remodeling of the ECM by the cell, we compare the orientation of ECM fibers at the start of the simulation to their orientation at later time points.
Specifically, for each fiber $i$, we define its center position $\vec{c}_i(t) = \vec{x}_{i, \lfloor N_\text{particles}/2\rfloor} - \vec{x}_\text{cell}$ relative to the cell's center of mass $\vec{x}_\text{cell}$ and the direction $\vec{v}_i(t)= \vec{x}_{i,\lfloor N_\text{particles}/2\rfloor+1} - \vec{x}_{i,\lfloor N_\text{particles}/2\rfloor}$ of the fiber's middle bond as in Figure~\ref{fig:remodelling}(A). 
We then compute the alignment $\gamma_i(t)$ of fiber $i$ along the cell as 
\begin{equation}
    \gamma_i(t) = \frac{\|\vec{c}_i(t)\| \|\vec{v}_i(t)\| } {\left| \vec{c}_i(t) \cdot \vec{v}_i(t)  \right|}.
\end{equation}
If the fibers are randomly oriented with respect to the cell, the values of \( \gamma_i(t) \) will be uniformly distributed on the interval \([0, 1]\). 
In contrast, a distribution of \( \gamma_i(t) \) values biased toward 1 indicates alignment of fibers pointing toward the cell.
To quantify this bias, we compute
\[
q(t) = \text{Prob}(0.9 \leq \gamma_i(t) \leq 1),
\]
and define the normalized alignment ratio as
\(
\frac{q(t)}{q_0},
\)
where \( q_0 = q(0) \) is the initial value. 
A value of \( q(t)/q_0 = 1 \) indicates no remodeling, while \( q(t)/q_0 > 1 \) reflects reorientation of fibers toward the cell. 

The quantity $q/q_0$ can be computed as a function of the distance from the cell $r$. In this case, the distribution of the inner product is computed as described above, with the exception that only fibers $i$ are considered where $\vec{c}_i(0) \leq r$. In this case, the fraction is determined $q(r)/q_0(r)$ at the final timepoint of the simulation.

\subsubsection{Densification factor}
The densification factor $\rho_\text{dens}$ is a measure of the densification of ECM around the cell by comparing the ECM density close to the cell with ECM density far away from the cell~\citep{malandrinoDynamic2019,tsingosHybrid2023}. Intuitively, a higher densification factor indicates accumulation of ECM by the cell. The densification factor $\rho_\text{dens}$ is defined as the ratio between $\rho_\text{close}$ and $\rho_\text{far}$. Before the densities $\rho_\text{close}$ and $\rho_\text{far}$ are computed, the fibers are interpolated by $10$ subnodes. That is, for each fiber $i$ we compute the subnodes $\vec{x}_{i,j,k}$ with $j=1,\dots,N_\text{particles}-1$ and $k=0,\dots,10$ via
\begin{equation}
    \vec{x}_{i,j,k} = \vec{x}_{i,j-1}\frac{k}{10} + \vec{x}_{i,j}\frac{10-k}{10}.
\end{equation}
Then, we construct two volumes, the first volume $V_\text{close}$ is that close to the cell and consists of all points with a distance at most $30\,\mu$m from the cell. The second volume $V_\text{far}$ is defined as the volume with distance at most $50\,\mu$m from the domain boundary. The densities are then computed as 
\begin{equation}
\rho_\text{close} = \frac{\#(\{x_{i,j,k}\}\cap V_\text{close})}{\text{volume}(V_\text{close})},
\end{equation}
and similarly for $\rho_\text{far}$ by replacing the subscripts. Finally the density factor is computed as $\rho_\text{dens} =\rho_\text{close}/\rho_\text{far}$. 

\section{Results}

\subsection{ECM elasticity limits cellular contraction}
\begin{figure}[h!]
    \centering
    \includegraphics[width=\linewidth]{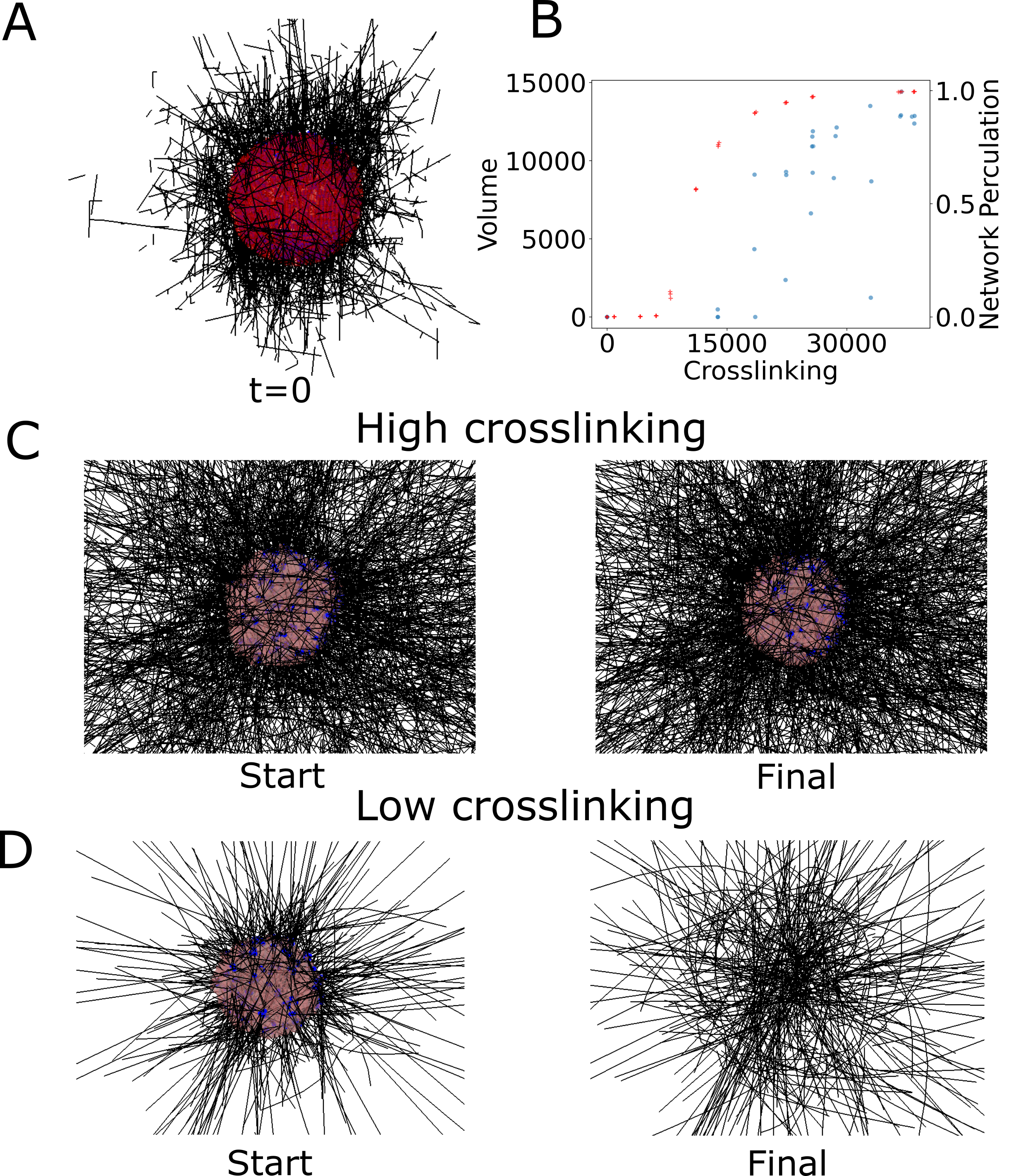}
    \caption{Contractile cell in ECM. A: Screenshot of the initial simulation state. B: A quantification of final cell volume as a function of ECM crosslinking in blue and the corresponding network percolation, the network's giant component, in red. C-D: Screenshot of simulations with the same crosslinking, $25000$ crosslinks, only those fibers drawn that moved at least $5\,\mu$m. C: Simulation with high crosslinking. D: Simulation with low crosslinking. }
    \label{fig:crosslink}
\end{figure}

The fundamental mechanism that we generalize from the 2D model~\citep{tsingosHybrid2023} to the 3D case is that the contractile forces of a cell can be countered by elastic restoring forces from the deformed ECM. 
    We show this by putting a completely contractile cell, i.e., a cell with target volume $V=0$, in a deformable ECM (Figure~\ref{fig:crosslink}(A)). In the absence of adhesion to the ECM, the cell contracts completely, as this is the energetically most favorable state (Equation~\eqref{eq:HamiltonH}). If the cell is initialized with adhesion to the ECM, the cell initially contracts and, as the cellular adhesions move, the cell deforms the ECM. The elastic force in the ECM resists the deformation and counteracts the cell's contraction. The balance between the cellular contractile and ECM-derived restoring forces can result in an equilibrium where the cell has positive volume (Figure~\ref{fig:crosslink}(B, C)). The mechanical equilibrium between cellular contraction and ECM resistance is governed by both the cell’s contractile force and the stiffness of the ECM. In bead-and-spring models of fibrous ECMs, this stiffness is primarily determined by the crosslink density~\citep{broederszModeling2014, tsingosHybrid2023}. Indeed, if we remove the crosslinkers from the network the cell deforms only the fibers to which it is initially attached. In this situation, there is no effective restoring force and the cell can completely retract (Figure~\ref{fig:crosslink}(D)). Section~\ref{sec:drag} provides a further analysis of this spurious drag and discusses how to keep it under control.

We first compared a non-crosslinked and a fully crosslinked ECM, which represents two extreme cases. The non-crosslinked ECM does not form a network as it is ill-connected, or not connected at all and does not mechanically resist deformation, where the crosslinked ECM does form a connected network that has elastic properties.  To test if the ECM stiffness and not only the ECM connectedness explain the difference in mechanical equilibrium, we vary $N_\text{cross,max}$ and measure the resulting cell volume. To ensure that the tested networks are sufficiently connected, we quantified their percolation using the giant component. The fiber network was treated as a graph, with particles as nodes and bonds or crosslinks as edges. The giant component of a graph with $n$ nodes is defined as $n_\text{connected}/n$, where $n_\text{connected}$ is the number of nodes in the largest connected component. Figure~\ref{fig:crosslink}(B) shows both the giant component and the final cell volume as a function of $N_\text{cross,max}$. The giant component increases sharply around 15,000 crosslinkers, indicating the emergence of a percolated network. Beyond this threshold, the equilibrium cell volume continues to increase with higher crosslinking. We therefore conclude that network stiffness, controlled by degree of crosslinking, determines the point at which the cell’s contractile force is balanced by the restoring force of the ECM, consistent with the 2D model~\citep{tsingosHybrid2023}.

\subsection{Fiber stiffness introduces effective drag on contraction}
\label{sec:drag}
As demonstrated above, restoring forces from the ECM resist the contractile forces generated by the CPM cell. This resistance is reflected by the equilibrium cell volume, which depends on the crosslinking between individual fibers. Here, we investigate whether the stiffness of individual fibers also influences this equilibrium.

\begin{figure}
    \centering
    \includegraphics[width=\linewidth]{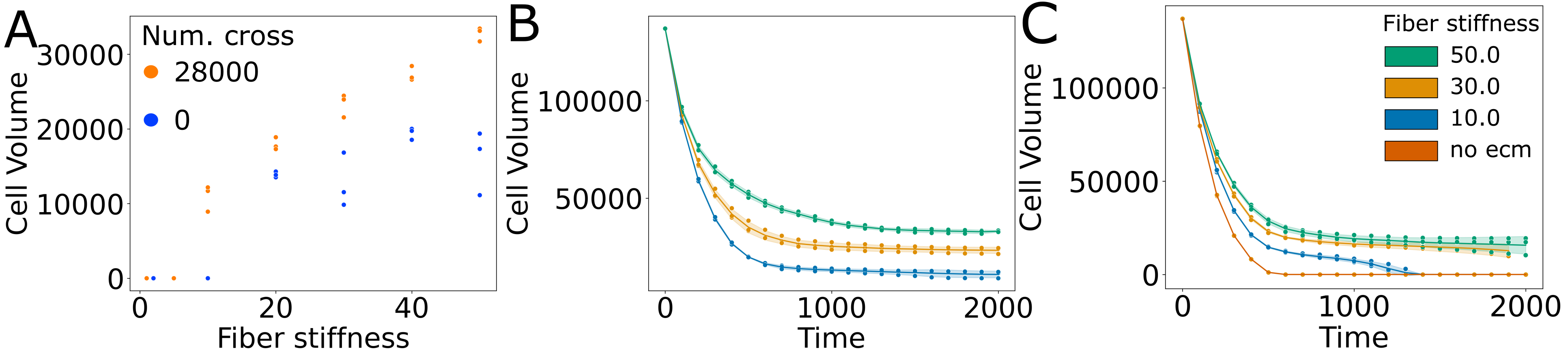}
    \caption{A: Dependence of fiber stiffness on equilibrium cell volume. Orange dots show simulations with 28000 crosslinkers and blue dots show simulations without crosslinkers. B: Cell volume as a function of time with crosslinkers. C: Cell volume as a function of time without crosslinkers.}
    \label{fig:drag}
\end{figure}

Figure~\ref{fig:drag}(A,B) show how the equilibrium cell volume in a crosslinked ECM increases with fiber stiffness. This can be understood by considering the ECM resistance experienced by the cell (Equation~\eqref{eq:deltah:ecm}), which scales linearly with the fiber stiffness $k$. However, Figure~\ref{fig:drag}(A) also shows an intriguing result for the non-crosslinked ECM (light-colored dots): even without crosslinks, the equilibrium cell volume remains non-zero when fiber stiffness is high.

This behavior can be explained by considering how the CPM and ECM are coupled, specifically through the movement of adhesion particles on fibers not crosslinked to other fibers. When an adhesion particle moves, the cell performs mechanical work (Equation~\eqref{eq:deltah:ecm}) deforming the associated fiber. After each MCS, the deformed fiber straightens again while the adhesion particle remains fixed, effectively shifting the fiber's position. In such scenarios, the ECM does not accumulate deformation, and subsequent movements of the same particle do not lead to progressively larger contributions from Equation~\eqref{eq:deltah:ecm}, because the fiber returns to a mechanically relaxed state after each step. Nonetheless, the deformation and subsequent relaxation of free fibers introduce a drag force. This drag can be observed clearly in the evolution of the cell volume within a non-crosslinked ECM, as shown in Figure~\ref{fig:drag}(C). Simulations with intermediate fiber stiffness ($k=10$) eventually reach a cell volume of $0$, similar to simulations without any ECM, but the contraction takes longer, confirming the existence of this drag force.

For simulations with higher stiffness values, as illustrated in Figure~\ref{fig:drag}(A,C), the cell maintains a positive equilibrium volume even without crosslinking. Here, the drag force exceeds the cell's contractile force, pointing to an unrealistic parameter regime.

Thus, two important caveats arise when considering the fiber stiffness parameter $k$. The first is the drag force itself, which influences the required duration (number of timesteps) of the simulation. The second caveat is that the drag force can surpass the cellular contractile force entirely. In such cases, the contributions from Equation~\eqref{eq:deltah:ecm} remain too large for the cell to contract, even when fibers are unconstrained. Thus, this parameter regime is unrealistic for modeling a deformable ECM and should therefore be avoided. In the remainder of this paper, we use parameter values around $k=10$, ensuring that this artifact has negligible influence. 

\subsection{ECM remodeling}
Thus far we have considered the effect of the ECM on the cell. However, the ECM structure is likewise affected by the cell's contraction. Here we consider the effect of cell contraction on fiber orientation and on ECM densification.

\subsubsection{Cell contraction reorients ECM}
After observing that cell contraction is countered by ECM deformation, we investigated whether and how the ECM structure changes in response to cell contraction. To quantify this, we consider the orientation of fibers relative to the cell at the beginning of the simulation, $q_0$, and at time $t$, $q(t)$, as shown in Figure~\ref{fig:remodelling}(A). The ratio $q(t)/q_0$ describes fiber orientation, where a value close to $1$ indicates randomly oriented fibers, and $q(t)/q_0>1$  indicates fibers pointing towards the cell.
We compared $q(t)/q_0$ across simulations with different levels of crosslinking to see how crosslinking affects ECM reorientation.

Figure~\ref{fig:remodelling}(B) shows the time evolution of $q(t)/q_0$, which increases with the number of crosslinkers. Clearly, fiber reorientation is strongest when the ECM is well-connected because the matrix displacement propagates further into the network, reorienting more fibers.

To examine spatial variations in fiber reorientation, we measured how the fraction $q/q_0$ depends on the distance $r$ from the cell boundary. Specifically, we analyzed the final simulation state $q(t_\text{end})$, computing $q$ and $q_0$ only for fibers located at distance $r$. Far from the cell, highly crosslinked ECM reorients more fibers, resulting in higher values of $q(r)/q_0(r)$ compared to ECM with lower crosslinking (Figure~\ref{fig:remodelling}C).

In networks with few or no crosslinkers, $q(r)/q_0(r)$ shows a small peak near the cell, caused by the reorientation of fibers initially adhered to the cell. In networks with more crosslinkers, a more pronounced peak emerges close to the cell boundary, showing that more nearby fibers reorient. However, this effect diminishes quickly with distance, and $q(r)/q_0(r)$ drops off accordingly. In contrast, highly crosslinked ECM networks show a less prominent peak near the cell, as crosslinking restricts local fiber reorientation. Nonetheless, the presence of crosslinks enables force transmission over longer distances, resulting in noticeable fiber alignment farther from the cell. These trends are consistent with prior observations that strain, whether externally applied or generated by cells, aligns collagen fibers and facilitates long-range force transmission \citep{vaderStrainInduced2009,pakshirDynamic2019}.

\subsubsection{ECM densification}
\begin{figure}
    \centering
    \includegraphics[width=\linewidth]{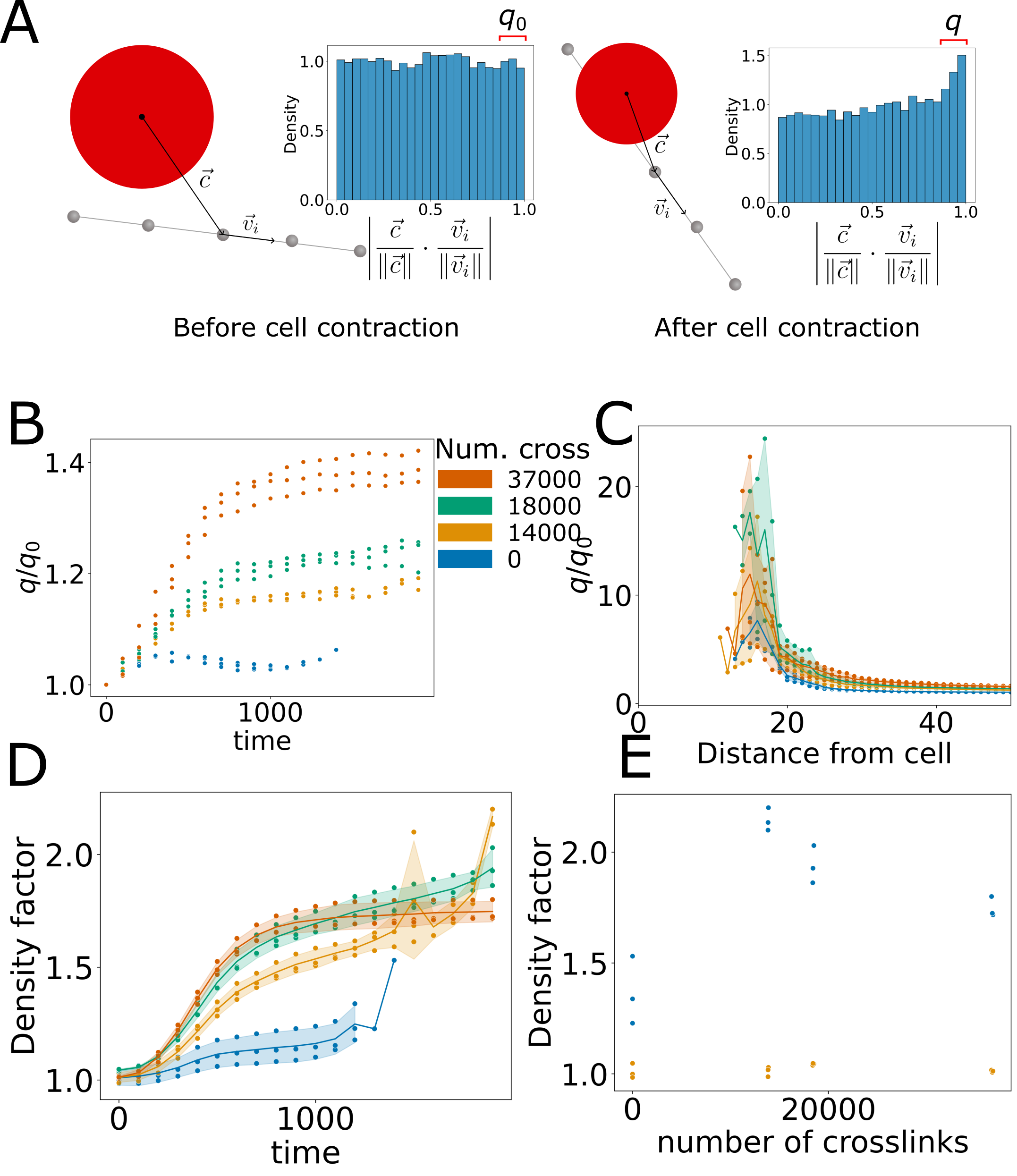}
    \caption{Contracting cells deform their surrounding matrix in ways that depend on ECM crosslinking.
(A) Graphical explanation of the fiber orientation measure, as described in the Methods section.
(B) Time evolution of the global orientation of the network. Increased crosslinking allows more fibers to be affected by cellular contraction, resulting in higher orientation values.
(C) Cumulative orientation fraction, computed using fibers located near the cell. Initial fluctuations reflect the small number of nearby fibers. A biphasic trend is observed: in non-crosslinked ECM, orientation remains low since only fibers initially attached to the cell are reoriented. In moderately crosslinked networks, a sharp peak appears as nearby fibers reorient, followed by a rapid decay because contraction does not propagate far. In highly crosslinked networks, the initial peak is lower due to restricted movement, but the decay is slower as the contraction propagates further and reorients more distant fibers.
(D–E) ECM density factor $\rho_\text{dens}$, indicating how the network compacts around the cell.
(D) Time series of network densification.
(E) Final densification values. For poorly connected networks in which the final cell volume reached zero, densification was measured just prior to collapse.}
    \label{fig:remodelling}
\end{figure}

Next we will study the accumulation of ECM around the cell, an effect known as ECM densification~\citep{malandrinoDynamic2019}. This behavior was previously observed in a 2D model~\citep{tsingosHybrid2023}. Here, we investigate whether the same behavior is reproduced in our 3D model.
We quantify ECM densification using the densification factor $\rho_\text{dens} = \rho_\text{close} / \rho_\text{far}$, where $\rho_\text{close}$ is the average ECM density close to the cell and $\rho_\text{far}$ the average density far from the cell. For details on how this is computed, we refer to the Methods section. In this section, we examine the relationship between $\rho_\text{dens}$ and ECM crosslinking.
We begin with a non-crosslinked network. In this case, the cell can only pull fibers that it is directly attached to, resulting in a low $\rho_\text{dens}$ (Figure~\ref{fig:remodelling}D). As crosslinking increases, the cell can recruit more fibers toward itself, as its displacements propagate further through the network. However, when crosslinking becomes too high, the densification factor decreases again (Figure~\ref{fig:remodelling}E). In this regime, the network is so rigid that contractile forces are more evenly distributed, reducing local fiber accumulation near the cell.

This biphasic relationship between $\rho_\text{dens}$ and crosslinking has been reported experimentally~\citep{malandrinoDynamic2019} and in the 2D model~\citep{tsingosHybrid2023}. Here, we show that our 3D model reproduces this behavior as well.

\section{Discussion}
In this paper, we extended the hybrid CPM of a cell in a fibrous ECM into 3D~\citep{tsingosHybrid2023}. This extension allows the exploration of the effect of cell contraction in 3D matrices. 

This development complements earlier computational studies, which often employed simplified ECM representations or were restricted to two-dimensional systems~\citep{oersMechanical2014,rensCell2020,tsingosHybrid2023,keijzerHow2025}. By capturing the discrete mechanical interactions between cells and ECM fibers, this approach offers new opportunities for detailed studies of multicellular dynamics in mechanically realistic 3D environments. 

For example, angiogenesis is an inherently 3D process, in which ECM plays an important role~\citep{duThreeDimensional2016,labelleSpatial2023,senkCollagen2021}. Extending the CPM-ECM coupling into 3D enables the development of more detailed models of angiogenesis that take the ECM mechanics into account. This could be used to study the role of fiber alignment and ECM stiffness during angiogenesis, factors which are relevant during tumor angiogenesis~\citep{balciogluTumorinduced2016}.

Moreover, this framework enables detailed studies of cell migration, where paths are shaped by mechanical heterogeneities, stiffness gradients, and active ECM remodeling by migrating cells, all of which can be explicitly modeled within the CPM. For example, a 3D CPM study found that ECM pore size strongly influences migration efficiency through nuclear constraints~\citep{sciannaCellular2021}. The here presented model could extend such analyses by allowing the ECM to remodel dynamically in response to cell-generated contractile forces, thereby capturing the feedback between cells and matrix.

Despite these advantages, several limitations remain. The current implementation uses static adhesion sites rather than dynamic, mechanosensitive focal adhesions, which strongly influence cell mechanics and ECM remodeling~\citep{mavrakisCompass2023}. Incorporating dynamic focal-adhesion models, as implemented in our 2D system~\citep{keijzerHow2025}, is a logical next step. In addition, we modeled a purely contractile cell to showcase how coupling to the ECM establishes a force balance between cellular contraction and ECM-derived restoring forces. Biological cells are not purely contractile, so effective contractility can be controlled via the target volume  $V$. Increasing $V$ allows for more flexible cell shapes and a wider range of dynamic behaviors.

Additionally, integrating this modeling technique into established simulation environments like Morpheus~\citep{starrussMorpheus2014} and CompuCell3D~\citep{swatMultiScale2012} would significantly improve its accessibility and usability. Such integration would facilitate adoption by the broader computational biology community, fostering collaborative development.

In summary, the presented framework broadens the scope of CPM-based approaches by enabling the study of cell–ECM interactions in mechanically detailed, three-dimensional environments. Addressing these limitations and enhancing integration into existing simulation toolchains will further increase its applicability and complement other modeling approaches in computational biology.

\section{Acknowledgments}
This publication is part of the ‘Mathematics-based strategies for repairing tumour blood vessels’ project with project number 865.17.004 to RM of the Vici research program, which is financed by the Dutch Research Council (https://www.nwo.nl/). The funders had no role in study design, data collection and analysis, decision to publish, or preparation of the manuscript. This work was performed using the compute resources from the Academic Leiden Interdisciplinary Cluster Environment (ALICE) provided by Leiden University.

\bibliography{sn-bibliography}

\end{document}